%% file: main.tex
\documentclass[conference]{IEEEtran}

\IEEEoverridecommandlockouts

\usepackage{comment}

\usepackage{booktabs}
\usepackage[utf8]{inputenc}
\usepackage[ruled,vlined]{algorithm2e}
\usepackage{microtype}
\usepackage{graphicx}
\usepackage{paralist}
\usepackage{tabularx}
\usepackage{balance}
\usepackage{multicol}
\usepackage{multirow}
\usepackage{pbox}
\usepackage{enumitem}
\usepackage{colortbl}
\usepackage{pifont}
\usepackage{xspace}
\usepackage{url}
\usepackage{tikz}
\usepackage{rotating}
\usepackage{balance}
\usepackage{float}
\usepackage[TABBOTCAP]{subfigure}
\usepackage{tcolorbox}
\usepackage{stackengine}
\usepackage{paralist}
\makeatletter
\newcommand{\linebreakand}{%
  \end{@IEEEauthorhalign}
  \hfill\mbox{}\par
  \mbox{}\hfill\begin{@IEEEauthorhalign}
}
\makeatother

\def\BibTeX{{\rm B\kern-.05em{\sc i\kern-.025em b}\kern-.08em
    T\kern-.1667em\lower.7ex\hbox{E}\kern-.125emX}}

\usepackage{listings}

\input{macro}

\begin{document}

\title{Do LLMs Provide Links to Code Similar \\ to what they Generate?\\  A Study with Gemini and Bing CoPilot}

\author{
\IEEEauthorblockN{Daniele Bifolco}
\IEEEauthorblockA{\textit{University of Sannio}\\
Benevento, Italy \\
d.bifolco@studenti.unisannio.it}
\and
\IEEEauthorblockN{Pietro Cassieri}
\IEEEauthorblockA{\textit{University of Salerno}\\
Fisciano, Italy \\
pcassieri@unisa.it}
\and
\IEEEauthorblockN{Giuseppe Scanniello}
\IEEEauthorblockA{\textit{University of Salerno}\\
Fisciano, Italy \\
gscanniello@unisa.it}
\and
\linebreakand %
\IEEEauthorblockN{Massimiliano Di Penta}
\IEEEauthorblockA{\textit{University of Sannio}\\
Benevento, Italy \\
dipenta@unisannio.it}
\and
\IEEEauthorblockN{Fiorella Zampetti}
\IEEEauthorblockA{\textit{University of Sannio}\\
Benevento, Italy \\
fzampetti@unisannio.it}
}


\maketitle
\thispagestyle{empty}

\begin{abstract}
Large Language Models (LLMs) are currently used for various software development tasks, including generating code snippets to solve specific problems. Unlike reuse from the Web, LLMs are limited in providing provenance information about the generated code, which may have important trustworthiness and legal consequences. While LLM-based assistants may provide external links that are ``related'' to the generated code, we do not know how relevant such links are.
This paper presents the findings of an empirical study assessing the extent to which 243 and 194 code snippets, across six programming languages, generated by Bing CoPilot and Google Gemini, likely originate from the links
provided by these two LLM-based assistants. The study leverages automated code similarity assessments with thorough manual analysis. 
The study's findings indicate that the LLM-based assistants provide a mix of relevant and irrelevant 
links having a different nature.
 Specifically, although 66\% of the links from Bing CoPilot and 28\% from Google Gemini are relevant, LLMs-based assistants still suffer from serious ``provenance debt''.
\end{abstract}

\begin{IEEEkeywords}
Large Language Models; Code Provenance; Empirical Study; Licensing; Trustworthiness
\end{IEEEkeywords}



\section{Introduction}
\label{sec:intro}
\input{intro}

\section{Study Design}
\label{sec:study}
\input{study}

\section{Study Results}
\label{sec:results}
\input{results}

\section{Distilled Findings and Implications}
\label{sec:implications}
\input{implications}

\section{Threats to Validity}
\label{sec:threats}
\input{threats}

\section{Related Work}
\label{sec:related}
\input{related}

\section{Conclusion}
\label{sec:conc}
\input{conclusion}

\section*{Data Availability}
\revised{The study replication package is available online~\cite{replication}.}

\section*{Acknowledgments}
\label{sec:ack}
\revised{This project has been financially supported by the European Union
NEXTGenerationEU project and by the Italian Ministry of the University and Research MUR, a Research Projects of Significant National Interest (PRIN) 2022 PNRR, project n. D53D23017310001
entitled ‘Mining Software Repositories for enhanced Software Bills
of Materials (MSR4SBOM)’. Daniele Bifolco is partially funded by
the PNRR DM 118/2023 Italian Grant for Ph.D. scholarships. Fiorella Zampetti is partially supported by funding from DM 1062/2021.}

\bibliographystyle{IEEEtranS}
\balance
\bibliography{bib.bib}

\end{document}

%% file: intro.tex
The availability of Large Language Models (LLMs) has profoundly impacted everyone's attitude to perform several tasks. This includes software developers that leverage LLMs for multiple activities, such as code generation, review, testing, repair, or documentation~\cite{fan2023large,hou2023large,msr2024}.

Focusing on leveraging LLMs for code generation, developers may be tempted to believe that asking an LLM to generate some code snippets (hereafter referred to as snippets) is not much different from reusing code from the Web. However, there are substantial differences, the first being the level of trust we can gain from the source code origin. When someone reuses code taken from an open-source project or even from a discussion in a Question \& Answer (Q\&A) forum, one can still leverage the authors' reputation/badges or posts' upvotes as proxies for trust. 
When we prompt an LLM for code generation, we rely solely on its reputation, size, and recent updates.
A second, but not less important, point concerns the legal implications of reusing and redistributing somebody else's code. Although LLMs are trained from artifacts mined from the Web, they take ownership of the generated code (\ie \textit{``AI-generated code. Review and use carefully''}). 
However, the generated code may be highly ``similar'' to some licensed code, and it is therefore important 
to determine the terms under which it can be reused and~redistributed.

Outside software development, 
LLM-generated artifacts have already triggered lawsuits between authors of newspaper articles or artistic creations and IT companies providing LLMs~\cite{karamolegkou-etal-2023-copyright}. Moreover, it was found that often LLMs fail to provide outputs accompanied by verifiable citations \cite{LiuZL23}.
Focusing on software development, recent studies have shown that LLMs, under specific prompts, can reproduce copyrighted source code~\cite{KatzyPDI24}. 
Summarizing, trustworthiness and legal requirements trigger the need to determine the provenance of a snippet generated by an LLM. At the time our research was conducted, most LLMs did not provide any information or valid licensing details for redistributing the generated code. Only a few LLM-based assistants provide the user with external links (hereafter referred to as links) related to the submitted query, especially if asked. This was the case of \bcp~\cite{copilot} and  \gem~\cite{geminiteam2023gemini}.
According to a recent survey \cite{li2023surveylargelanguagemodels}, LLM could use three different strategies to provide attribution, namely (i) direct model-driven, (ii) post-generation answering, and (iii) post-generation. So far, the first option has not been utilized (as it often leads to hallucination), while the second mainly works for Retrieval Augmented Generation (RAG) engines. The third, which uses a post-retrieval to attribute sources, is the one used by \bcp and \gem.

To date, no study evaluated the usefulness of the provided links in determining the generated code provenance.
To fill in this gap, we report the results of an empirical study aimed at investigating the following general research question:
\begin{quote}
\emph{To what extent can LLM-based assistants provide reliable provenance information for the source code they generate?}
\end{quote}

In the context of our study, we consider two popular LLM-based assistants---\bcp~\cite{copilot} and \gem~\cite{geminiteam2023gemini}---different in their architecture, both able to provide external links related to the submitted query. For code generation tasks, we use the \textsc{CodeSearchNet} dataset~\cite{husain2020codesearchnetchallengeevaluatingstate}, which provides realistic natural language queries related to coding tasks. For each coding task, the dataset also provides at least one code snippet solving it with  \lang programming languages: Go, Java, JavaScript, PHP, Python, and Ruby. 
We first execute the queries against the two assistants, using a sequence of prompts to ask for solving the coding tasks and provide links related to them. Then, we combine (i)~an automated analysis in which we scrape snippets from the provided links and run a clone detection tool (\textsc{CCFinderSW}~\cite{8305997}) and textual similarity between the LLMs-generated snippets and the ones retrieved from the links; and (ii)~an extensive manual analysis over 1,520 links to categorize the link types, analyze their content, and determine whether the link is relevant to the query and contains code snippets likely matching the generated~ones. 

The paper's contributions can be summarized as follows:
\begin{itemize}

\item We report the findings of an empirical study on the ability of two LLM-based assistants (\ie \bcp and \gem) to provide reliable provenance information about their generated code and discuss implications and challenges for developers, researchers, and LLM~providers;
\item We provide a dataset featuring manually validated LLM results, as well as results from automated clone detection and textual similarity on the generated snippets~\cite{replication}. The dataset comprises \cNTqueries queries, \cSnip snippets, and \cSources links for \bcp, and \gNTqueries queries, \gSnip snippets and \gSources links for \gem.
\end{itemize}

\revised{The paper is organized as follows. \secref{sec:study} describes the study definition, research questions, and planning. Results are reported in \secref{sec:results}, while their implications are discussed in \secref{sec:implications} and their threats to validity in \secref{sec:threats}. After a discussion of related literature about code provenance and code search (\secref{sec:related}), \secref{sec:conc} concludes the paper and outlines directions for future work.}

%% file: study.tex
The \emph{goal} of our study is to investigate the extent to which LLM-based assistants, when used for solving coding tasks, can provide reliable provenance information for the generated code. 
The \emph{context} consists of two popular LLM-based assistants, \ie \bcp ~\cite{copilot} and \gem~\cite{geminiteam2023gemini}, and 99 natural language queries representative of coding tasks, for six programming languages, \ie Go, Java, JavaScript, PHP, Python, and Ruby, extracted from a manually curated benchmark~\cite{husain2020codesearchnetchallengeevaluatingstate}.  

\begin{figure}[t!]
    \centering
    \includegraphics[width=0.8\linewidth]{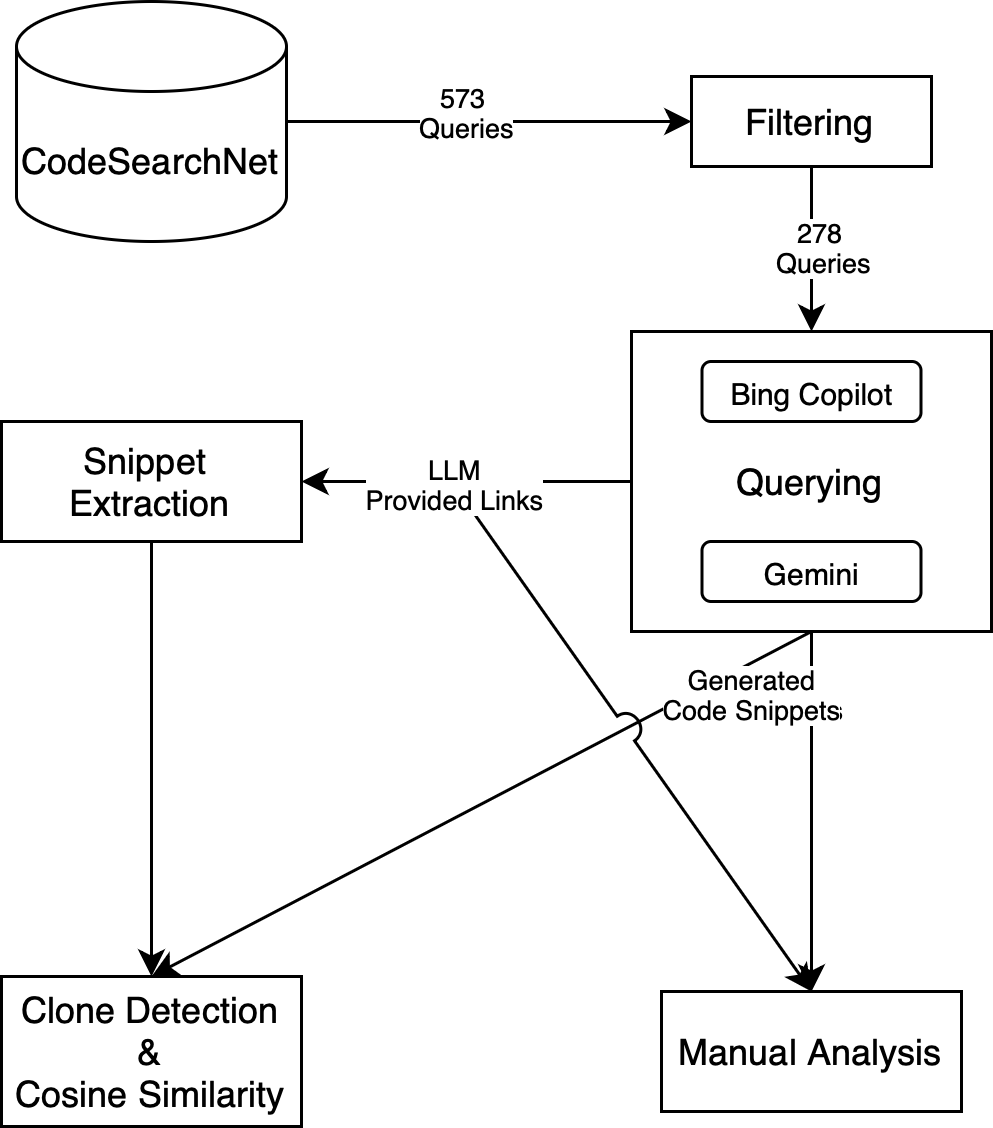}
    \caption{Dataflow of the Study Design}
    \label{fig:dataflow}
\end{figure}

We aim to answer the following two research questions (RQs):
\begin{itemize}
    \item \textbf{RQ$_1$:} \emph{\rqone}
    \item \textbf{RQ$_2$:} \emph{\rqtwo}
\end{itemize}
First, we look at the number and types of links provided by the LLM-based assistants, along with the generated snippets. Then, we investigate whether snippets contained in the link landing page likely represent the provenance of the LLM-generated snippet.

\revised{The study methodology is shown in \figref{fig:dataflow}. First, we select relevant queries from the CodeSearchNet dataset. After that, on the one side, we run the queries against \bcp and \gem. From the query results, we perform a manual analysis to categorize the provided links and to determine the extent to which the code shown in those links is similar to the one generated by the LLM. On the other side, we complement this manual analysis with a clone detection/code similarity analysis between snippets extracted from the landing pages of the links and those generated by the LLMs.}


\subsection{Study Context}


\textbf{Choice of LLM-based assistants.}
Although we do not aim to generalize the results to all code-related LLMs, the selection of the LLM-based assistants used in this study warrants discussion. The selection procedure was made by accounting for the LLM's ability to provide external links related to a query, which was possible for both \bcp and \gem, LLM-based assistants providing links by leveraging post-generation attribution \cite{li2023surveylargelanguagemodels}.
Our analysis highlighted how ``open'' models, such as Llama-3~\cite{llama3} or its code-specific counterpart, \ie Code Llama~\cite{abs-2308-12950}, do not provide additional information as links. The same applies to GitHub Copilot~\cite{github_copilot}, a widely-used IDE-integrated tool \revised{which chat is} based on GPT-4. For ChatGPT, on top of GPT-3.5 (used by ChatGPT when we conducted the study), we could not obtain any links, even when we directly asked for them. Only for \bcp~\cite{copilot} and \gem~\cite{geminiteam2023gemini}, we could obtain links likely containing snippets related to the LLM-generated snippets. 

\textbf{Dataset.}
To achieve our goal, we needed to collect a set of natural language queries describing coding tasks to use for querying the LLM for code generation. Instead of arbitrarily defining them, and thus limiting internal validity threats, we considered 99 queries representative of coding tasks implemented in six widely adopted programming languages, \ie Go, Java, JavaScript, PHP, Python, and Ruby, contained in a well-known and acknowledged benchmark, \ie \textsc{CodeSearchNet}~\cite{husain2020codesearchnetchallengeevaluatingstate}, already used to evaluate code search approaches, as well as the LLM effectiveness in code generation tasks~\cite{vaithilingam2022expectation, wang2023codet5+,xu2022systematic}. It is worth noting that \textsc{CodeSearchNet} includes realistic implementation tasks, such as “aes encryption”, “linear regression” or “parse command line arguments.”

\begin{table}[t!]
\caption{Dataset Characterization, \ie number of queries (coding tasks) by programming language}
\label{tab:dataset}
\centering
    \begin{tabular}{lrrrr}
    \toprule
    \multirow{2}{*}{\textbf{Language}} & \multirow{2}{*}{\textbf{Queries~\cite{husain2020codesearchnetchallengeevaluatingstate}}} & \textbf{Exact} & \multicolumn{2}{c}{\textbf{No trivial}} \\&&\textbf{Match~\cite{husain2020codesearchnetchallengeevaluatingstate}}&\textbf{\bcp} & \textbf{\gem} \\
    \midrule
    \textbf{Go} & 83 & 8 & 7 & 7\\
    \textbf{Java} & 99 & 65 & 46 & 48\\
    \textbf{JavaScript} & 96 & 44 & 28 & 30\\
    \textbf{PHP} & 99 & 43 & 25 & 25\\
    \textbf{Python} & 99 & 92 & 55 & 48\\
    \textbf{Ruby} & 97 & 26 & 14 & 15 \\
    \midrule
    \textbf{Overall} & 573 & 278 & 175 & 173 \\
    \bottomrule
    \end{tabular}
    \vspace{-2mm}
\end{table}

As shown in \tabref{tab:dataset}, we started looking at the original set of 573 queries \revised{(the sum is not 594 because some queries were not present in some languages)} in the dataset and retained only those for which the \textsc{CodeSearchNet} authors retrieved at least one snippet on the Web addressing them.  The rationale of this filtering is to restrict the set of queries only to those for which we know that there exists a snippet on the Web that solves the coding task. This is because, if we are sure the snippet is available on the Web, the LLM could be able to satisfy the query by simply reproducing it.
By applying the aforementioned filtering, we ended up with 278 queries spread differently across programming languages (see \tabref{tab:dataset}), \eg we have only eight queries for Go, while for Python we kept 92 queries. Note that the 278 queries consist of instances of the 99 initial ones on the \lang programming languages, pending the filtering described above.

\subsection{Step 1: Querying the LLM (and search engine)}
\label{sec:querying}
To study our RQs, we prompted LLM-based assistants to provide code snippets addressing a coding task in a specific programming language and related links where it is possible to find code snippets similar to the ones generated. 
After experimenting with three different prompts (available in our replication package~\cite{replication} and described below), we ended up with prompts that were more effective in providing both the snippet that solved the task and relevant links.

We first tried with a prompt containing the programming language and the query:
\begin{quote}
\emph{``In $<<$LANGUAGE$>>$ : $<<$QUERY$>>$"}
\end{quote}
However, by using this prompt on a random set of 56 queries for the two LLM-based assistants, both \bcp and \gem could not properly understand our request in 25 and 14 cases, respectively, \ie no code was provided.
Therefore, in the second attempt, we improved the query by specifying the LLM role~\cite{3613905.3650786} and clarifying the programming language to be used to perform the task: 

However, we found that simply putting the programming language near the role did not work, so in the third and final attempt, we had to repeat it twice:
\begin{quote}
\emph{``You are a Senior $<<$LANGUAGE$>>$ developer. Then give me a $<<$LANGUAGE$>>$ code snippet about: $<<$QUERY$>>$''}
\end{quote}
Note that no additional prompt engineering was required to enhance the LLM’s ability to return relevant code. This is because (i) a simple prompt was sufficient for the LLM to generate code satisfying the query, and (ii) we manually verified that the generated code snippet implemented what was specified in the query.

While \bcp directly provides as output both the code snippets and a set of links where it would be possible to find additional information about the query, for \gem there are cases (140 out of 278 queries) for which to obtain the links, we needed a follow-up prompt explicitly asking for: 

\begin{quote}
    \emph{``Where can I find more information about this code?"}
\end{quote}
This somewhat mimics what \bcp directly provides, telling where to ``Learn More''. 

We found that asking about ``more information" yielded better results than ``sources", making interactions with \gem more consistent with those using \bcp.
When prompted with ``Give me the sources", we often obtain responses like ``The code snippet I provided is written by me".
Lastly, we experimented with a chain-of-thought approach (see \secref{sec:threats}), but this did not yield any improvement in results.

The querying procedure was conducted manually. Specifically, for each query, we (1) cleaned up the LLM interaction session, (2) sent the prompt, and (3) saved the returning output, \ie generated snippets and links (if any). 
The exact prompts and the generated snippets with related links are available in our replication package~\cite{replication}.

Since we could not obtain links when interacting directly with the models, it is possible that \bcp and \gem perform a Web search through the respective search engines~\cite{ma2024crafting}, \ie Bing~\cite{bing} and Google~\cite{google}. 
To verify this conjecture, we queried the two search engines using the following prompt:
\begin{quote}
``$<<$LANGUAGE$>>$ code snippet about $<<$QUERY$>>$''  
\end{quote}
This prompt is slightly different, but consistent, from the one used with the assistants, as it better adheres to query the search engines. Using the same prompt used for the assistant on the search engine, we mainly obtained pages on how to become an experienced developer in a given programming language.

The interaction with search engines was automated using a Python script and AppleScript~\cite{cook2007applescript},
by extracting the top-$k$ returned links, where $k$ corresponds to the number of links returned by the LLM for the same query.

\subsection{Step 2: Manual Analysis}
\label{sec:manual}
We performed a manual analysis to evaluate the effectiveness of LLM-based assistants in providing links containing ``similar'' code snippets to the ones they generate. Specifically, for each assistant, we prepared a spreadsheet reporting the query, the automatically generated snippets, and the provided links (one per row).
Before starting the manual analysis, we checked whether the links provided by the LLMs overlap with those in the \textsc{CodeSearchNet}. We found no overlap, hence we avoid that the choice of the dataset could have biased our results.

Five authors of the paper (hereafter referred to as annotators) were involved in the process. The total set of queries provided by the two LLM-based assistants (\cNTqueries for \bcp and \gNTqueries for \gem) was evenly distributed among the five annotators, ensuring that two people independently analyzed both the automatically generated code and the provided links for each query.
The annotation procedure aimed at evaluating the following four dimensions. 

\textbf{Excluding trivial snippets.} For each generated snippet, the annotators had to determine whether it was trivial. They did this by evaluating the likelihood that a similar snippet could appear ``by chance'' on the Web, due to the lack of context-specific information and because this is essentially the only way such a task could be accomplished.
Annotators agreed in 80.4\% of the cases (607 out of 755 automatically generated snippets). To check for agreement by chance, we computed Cohen's $k$~\cite{cohen1960coefficient}, which was 0.58, indicating a moderate level of agreement among the annotators. Note that we limited issues due to agreement by chance because all cases, regardless of agreement or not, were double-checked during the discussion phase.
As reported in \tabref{tab:dataset}, we ended up with 175 and 173 queries having at least one non-trivial generated snippet for \bcp and \gem, respectively. The remaining analyses, as well as the overall results, have been conducted considering only the links belonging to these queries, \ie 1,520 links, of which 1,006 belong to \bcp and 514 to \gem. Note that, the number of non-trivial snippets for Go and Ruby is substantially lower than for other languages, thus their results must be interpreted cautiously. 

\textbf{Classifying the links.} The annotators characterized the type of links the assistant provides for a specific query. \textit{Not a Source} has been used to account for cases where the links did not contain any snippets, while \textit{Unreachable} refers to links that did not work.
The annotators classified the remaining links using a set of categories among those drafted in a pivotal round involving the links related to 10 queries submitted to both assistants. These categories are Q\&A, Repository, Tutorial, Official Documentation, Issue/Pull Request, and Gist/Code Example. Annotators could add further categories if needed. The annotation procedure introduced four new categories, \ie Video, Wiki, Sandbox, and PDF document. Disagreements among annotators occurred in only 160 out of \Sources links provided by the two assistants ($\simeq 10.5\%$). For this coding, we reached Cohen's $k$ of 0.91, \ie almost a perfect agreement.

\textbf{Classifying the link relevance.} The annotators checked whether or not the provided link was related to the query, choosing among three different options: ``yes", ``no" or ``related to the query but with a different programming language''.  
Disagreements among annotators only occurred in $\simeq 7.2\%$ of the cases (109 out of \Sources external links), resulting in Cohen's $k=0.83$,\ie almost a perfect agreement.

\textbf{Classifying relevance of the snippets in the links and the LLM-generated snippets.} As a last step, the annotators assessed the extent to which the generated snippets likely originate from the provided links by using a yes/no evaluation. This was done by looking not only at code similarity in terms of its structure but also, above all, at elements such as identifiers and literals that could be associated with a given Web link.
Disagreements occurred in 199 out of \Sources links ($\simeq 13.1\%$), resulting in Cohen's $k=0.77$, \ie a strong/substantial agreement. 

\subsection{Step 3: Automated Similarity Analysis}
To complement the manual analysis and understand whether clone/similarity analysis can support provenance analysis for LLM-generated code, we analyzed the similarity between snippets contained in the links and those generated by the assistants.
First, for each link, we automatically extracted the snippets for common domains with identifiable separators (\eg HTML tags), \eg text from Stack Overflow is pulled from the nested \texttt{pre} and \texttt{code} tags, while GitHub repositories are accessed via GitHub raw URLs. All other extractions are performed manually.
Further details can be found in the replication package~\cite{replication}. 

Once snippets have been extracted, we leveraged two types of automated analysis to determine the likely relevance of the snippets available in the links to the LLM-generated ones. First, we employed clone detection, which has already been used in the past for provenance analysis~\cite{GermanPGA09}, relying on \textsc{CCFinderSW}~\cite{8305997}, \ie an evolution of \textsc{CCFinder}~\cite{KamiyaKI02}. Such a tool leverages a token-based approach to identify Type-1 and Type-2 clones. The reasons for choosing this tool are its ability to (i) analyze all six considered programming languages and (ii) parse partial snippets. 
To avoid the detection of small clones, which may occur by chance, we calibrated the tool with a threshold of at least 20 tokens (using the tool's option \texttt{-t 20}), which corresponds, on average, to five lines of code. 
Choosing the proper threshold is quite controversial, and previous studies have set it to a size between 20 and 50 tokens~\cite{BellonKAKM07,KamiyaKI02, KimSN05, roy07,2884781.2884877}. To avoid missing many clones, and since we checked false positives by hand anyway, we chose a lower bound, in line with what was done by Bellon \etal~\cite{BellonKAKM07}.
The tool outputs clone pairs---satisfying the threshold specified above---between snippets and, for each clone pair, its length. Using this information, we computed a \emph{cloning ratio}, defined as clone length (in terms of non-empty lines) divided by the length of the LLM-generated snippets.

As \textsc{CCFinderSW} does not distinguish between Type-1 and Type-2 clones, one risk of the clone detection is that it may report Type-2 clones that, while similar to the LLM-generated code in terms of structure, are unlikely to be the training source for the LLM, \eg having quite different identifiers or literals. For this reason, we complemented the cloning ratio with a textual similarity. To this aim, we built a vector space model using the Python \emph{CountVectorizer}, considering each token as a vocabulary word (we leveraged the tokenization performed by \textsc{CCFinderSW}), and as weight the number of token occurrences in the snippet. To avoid obtaining high similarity between structurally similar snippets, we excluded the reserved words and separators (\eg ``\texttt{\{\}.;()}'' for Java) whose lists are available in \textsc{CCFinderSW}. At the same time, we kept identifiers and literals as distinct words and did not combine them into a single token.
After that, we computed the cosine similarity between the LLM-generated snippet and those retrieved from the link.

\subsection{Analysis Methodology}
\label{sec:method}

To answer RQ$_1$, we first report descriptive statistics 
of the number of links reported by the LLMs for the performed queries.  Then, we provide a breakout of the most frequent domains for the returned links, 
and the percentage of links relevant to the query. 
Lastly, we compare the links returned by the LLMs with those returned by the respective search engines. Specifically, given a query for which an LLM has returned $k$ links, we compute and report the Jaccard overlap~\cite{jaccard} (intersection/union ratio) between these links and the top-$k$ links reported by the search engine when performing the same~query.

\begin{table}[b]
\vspace{-2mm}
\caption{Descriptive statistics of the number of links to external sources provided by the LLM-based assistants}
\label{tab:sources}
\centering
\resizebox{0.8\columnwidth}{!}{
\begin{tabular}{lrrrrrr}
    \multicolumn{7}{c}{\sc Bing CoPilot} \\ 
    \midrule
    \textbf{Languages} & \textbf{Min} & \textbf{1Q} & \textbf{Median} & \textbf{Mean} & \textbf{3Q} & \textbf{Max} \\
    \midrule
        \textbf{Go} &   4 & 4.00 & 4.00 & 4.86 & 5.50 &   7 \\ 
        \textbf{Java} &   3 & 3.00 & 4.00 & 5.24 & 6.00 &  18 \\ 
        \textbf{JavaScript} &   4 & 4.00 & 5.50 & 6.04 & 7.00 &  12 \\ 
        \textbf{PHP} &   3 & 4.00 & 5.00 & 5.92 & 8.00 &  13 \\ 
        \textbf{Python} &   3 & 5.00 & 5.00 & 5.89 & 7.00 &  13 \\ 
        \textbf{Ruby} &   3 & 3.25 & 5.00 & 5.57 & 6.00 &  15 \\ 
   \midrule 
   \multicolumn{7}{c}{\sc Gemini} \\ 
   \midrule
   \textbf{Languages} & \textbf{Min} & \textbf{1Q} & \textbf{Median} & \textbf{Mean} & \textbf{3Q} & \textbf{Max} \\
   \midrule
        \textbf{Go} &   2 & 3.00 &   3.00 & 3.57 & 3.50 &   7  \\ 
        \textbf{Java} &   1 & 1.75 &   2.00 & 2.65 & 4.00 &   6 \\ 
        \textbf{JavaScript} &   1 & 2.00 &   3.00 & 3.43 & 4.00 &   8 \\ 
        \textbf{PHP} &   1 & 2.00 &   3.00 & 3.24 & 4.00 &   8 \\ 
        \textbf{Python} &   1 & 1.75 &   3.00 & 2.83 & 3.25 &   7 \\ 
        \textbf{Ruby} &   1 & 2.00 &   3.00 & 2.80 & 3.00 &   5 \\ 
   \bottomrule
\end{tabular}
}
\vspace{-1mm}
\end{table}

\begin{table*}[t]

\caption{Top-10 domains for the LLM-based assistants provided links (Q\&A: Question\&Answare, R: Repository, T: Tutorial, O: Official Documentation, Ex: gist/example, W: Wiki, V: Video, S: Sandbox, NA: Unreachable/Not a Source)}
\label{tab:topdomain}
\centering
\resizebox{\columnwidth*2}{!}{
\begin{tabular}{llllll}
    \multicolumn{6}{c}{\sc \bcp} \\ 
    \midrule
    \textbf{Go} & \textbf{Java} & \textbf{JavaScript} & \textbf{PHP} & \textbf{Python} & \textbf{Ruby} \\ 
    \midrule
        Q\&A: stackoverflow.com (9) & Q\&A: stackoverflow.com (58) & Q\&A: stackoverflow.com (44) & Q\&A: stackoverflow.com (36) & Q\&A: stackoverflow.com (102) & Q\&A: stackoverflow.com (33) \\ 
          R: github.com (7) & R: github.com (40) & R: github.com (29) & R: github.com (32) & NA: gettyimages.com (23) & O: ruby-doc.org (3)  \\ 
          W: wikipedia.org (3) & T: baeldung.com (25) & NA: gettyimages.com (20) & O: php.net (12) & T: geeksforgeeks.org (19)  & W: wikipedia.org (3) \\ 
          O: pkg.go.dev (3) & T: geeksforgeeks.org (22) & T: geeksforgeeks.org (10) & W: wikipedia.org (12) & R: github.com (15) & O: docs.ruby-lang.org (3) \\ 
          T: tutorialgateway.org (1) & NA: gettyimages.com (6) & O: developer.mozilla.org (6) & T: geeksforgeeks.org (3) & T: blog.finxter.com (8)  & O: rubydoc.info (3) \\ 
          T: codingster.com (1) & T: mkyong.com (5) & NA: cheeso.members.winisp.net(4) & T: w3schools.com (2) & V: youtube.com (7) & R: github.com (3) \\ 
          T: reintech.io (1) & O: docs.oracle.com (4) & T: freecodecamp.org (3) & T: dev.to (2) & T: pythonexamples.org (7)  & O: learn.microsoft.com (2) \\ 
          Ex: onelinerhub.com (1) & T: howtodoinjava.com (4) & T: w3schools.com (3) & T: codexworld.com (2) & W: wikipedia.org (5)  & T: w3schools.com (2) \\ 
          NA: echoof.me (1) & T: codeease.net (3) & W: wikipedia.org (3) & NA: example.org (2) & O: learn.microsoft.com (5) & T: rubyguides.com (2) \\ 
          Ex: golangcode.com (1) & O: commons.apache.org (3) & NA: onicos.com (2) & S: pastebin.com (2) & T: datagy.io (4) & NA: example.com (2) \\ 
    \midrule 
    \multicolumn{6}{c}{\sc \gem} \\ 
    \midrule
    \textbf{Go} & \textbf{Java} & \textbf{JavaScript} & \textbf{PHP} & \textbf{Python} & \textbf{Ruby} \\ 
    \midrule
        O: pkg.go.dev (6) & R: github.com (35) & O: developer.mozilla.org (27) & O: php.net (47) & O: docs.python.org (38) & Q\&A: stackoverflow.com (7) \\ 
          Q\&A: stackoverflow.com (4) & Q\&A: stackoverflow.com (25) & R: github.com (13) & Q\&A: stackoverflow.com (7) & Q\&A: stackoverflow.com (13) & O: docs.ruby-lang.org (7) \\ 
          O: go.dev (4) & O: docs.oracle.com (21) & Q\&A: stackoverflow.com (10) & R: github.com (6) & R: github.com (9) & R: github.com (6) \\ 
          W: wikipedia.org (3) & T: baeldung.com (11) & O: npmjs.com (7) & PDF: repositorio.ul.pt (2) & T: realpython.com (7) & O: ruby-doc.org (5) \\ 
          T: geeksforgeeks.org (2) & W: wikipedia.org (5) & T: w3schools.com (5) & NA: sourceforge.net (2) & T: geeksforgeeks.org (6) & T: rubyguides.com (3) \\ 
          T: tutorialspoint.com (1) & O: commons.apache.org (3) & W: wikipedia.org (4) & W: wikipedia.org (2) & O: numpy.org (5) & T: sitepoint.com (2) \\ 
          T: digitalocean.com (1) & T: javatpoint.com (3) & NA: plotly.com (4) & O: developer.mozilla.org (2) & T: w3schools.com (4) & O: rubydoc.info (2) \\ 
          V: youtube.com (1) & T: oracle.com (3) & T: medium.com (3) & T: symfony.com (1) & O: requests.readthedocs.io (4) & O: learn.microsoft.com (2) \\ 
          T: yourbasic.org (1) & T: geeksforgeeks.org (3) & T: themurmuroussea.tistory.com (3) & T: codementor.io (1) & O: python-reference.readthedocs.io (3) & W: wikipedia.org (2) \\ 
          NA: gonum.org (1) & NA: algs4.cs.princeton.edu (2) & T: codeyourfuture.github.io (3) & T: programmierfrage.com (1) & W: wikipedia.org (3) & NA: rubygems.org (1) \\ 
    \bottomrule
\end{tabular}
}
\vspace{-2mm}
\end{table*}  

To answer RQ$_2$, 
we report the percentage of queries for which (i) the LLM returned at least one snippet, (ii) where it returned more than one snippet considered relevant by the annotators, and (iii) the number of links, by category, containing snippets relevant to the LLM-generated ones. 
Finally, we compare the results of the manual analysis with those obtained through clone detection and textual similarity. 
As for the clone detection, we compute, for each reported link, the maximum cloning ratio between all snippets embedded in the link and the LLM-generated snippet. We consider the maximum because we are interested in whether there is at least one snippet in the link highly similar to the generated one.
We show the distribution of such similarities, comparing them between links confirmed by the annotators and links classified as ``no''. We complemented this analysis by the (non-parametric) Wilcoxon rank sum test~\cite{wilcoxon1992individual} (data were not normally distributed as the Wilk-Shapiro test indicated), and Cliff delta effect size~\cite{cliff1996answering}, considering a significance level of 95\%. Due to multiple tests, $p$-values are adjusted using Holm's correction~\cite{holm1979simple}.
A similar analysis is performed for the textual~similarity.

%% file: results.tex
In this section, we report the results of each defined RQ.

\begin{table}[t!]
\caption{Relevance of the links to the query}
\label{tab:related_to_queries}
\centering
\resizebox{0.9\columnwidth}{!}{
\begin{tabular}{lrrrr}
    \multicolumn{5}{c}{\sc \bcp} \\ 
    \midrule
    \multirow{2}{*}{\textbf{Language}} & \multirow{2}{*}{\textbf{Yes}} & \multirow{2}{*}{\textbf{No} }& \textbf{Related but} & \multirow{2}{*}{\textbf{NA}} \\ 
    & & & \textbf{different language} & \\
    \midrule
      \textbf{Go} & 58.82\% & 26.47\% & 2.94\% & 11.76\% \\ 
      \textbf{Java} & 61.79\% & 13.41\% & 6.50\% & 18.29\% \\ 
      \textbf{JavaScript} & 56.21\% & 7.69\% & 13.02\% & 23.08\% \\ 
      \textbf{PHP} & 56.76\% & 17.57\% & 10.14\% & 15.54\% \\ 
      \textbf{Python} & 54.98\% & 8.46\% & 9.97\% & 26.59\% \\ 
      \textbf{Ruby} & 41.03\% & 10.26\% & 23.08\% & 25.64\% \\ 
   \midrule 
   \multicolumn{5}{c}{\sc \gem} \\ 
   \midrule
    \multirow{2}{*}{\textbf{Language}} & \multirow{2}{*}{\textbf{Yes}} & \multirow{2}{*}{\textbf{No} }& \textbf{Related but} & \multirow{2}{*}{\textbf{NA}} \\ 
    & & & \textbf{different language} & \\
   \midrule
      \textbf{Go} & 56.00\% & 28.00\% & 4.00\% & 12.00\% \\ 
      \textbf{Java} & 29.13\% & 15.75\% & 4.72\% & 50.39\% \\ 
      \textbf{JavaScript} & 26.21\% & 36.89\% & 5.83\% & 31.07\% \\ 
      \textbf{PHP} & 48.15\% & 30.86\% & 0.00\% & 20.99\% \\ 
      \textbf{Python} & 24.26\% & 48.53\% & 2.94\% & 24.26\% \\ 
      \textbf{Ruby} & 42.86\% & 33.33\% & 2.38\% & 21.43\% \\ 
   \bottomrule
\end{tabular}}
\vspace{-3mm}
\end{table}

\subsection{RQ$_1$: \rqone}
\tabref{tab:sources} summarizes the distributions of the number of links provided for different programming languages by both \bcp and \gem. Recall that this only accounts for the queries (\cNTqueries for \bcp and \gNTqueries for \gem) for which at least one of the automatically-generated snippets was considered as non-trivial.
For \bcp, regardless of the programming language, the minimum number of provided links for a query is never less than three (median value in the range 4.0-5.5). For \gem, this number drops to one in five out of the six programming languages, with a median value of almost three, except for Java, which has a median value of two.
When \bcp was asked to provide a Java code snippet for \textit{httpclient post json}, it returned 18 links.
However, we found that only five contained Java code related to the query, and six were related to the query but with different programming languages. For the same query, \gem only provided three links, of which only one contains Java code related to the query. A similar behavior is observed when requesting a Ruby code snippet to \textit{parse query string in url}. In this case, \bcp provided 15 links, of which only three contained Ruby code relevant to the query, while 
\gem returned four links, with one being related to the query.

\tabref{tab:topdomain} lists the top-10 domains across all links provided for the queries, broken down by programming language, for \bcp and \gem. Each domain is preceded by its type and followed by the number of its occurrences. Due to space limits, a complete breakdown of link types by programming language and LLM-based assistant is available in our replication package~\cite{replication}.

For \bcp, regardless of the programming language, Q\&A (Stack Overflow) posts are the most frequently retrieved, followed by (R)epositories, \ie GitHub. 
\gem tends to provide links to (O)fficial documentation documentation, such as \textit{pkg.go.dev}, \textit{developer.mozilla.org}, \textit{php.net}, or \textit{docs.python.org}. 
As domains not containing code snippets, \gem only returns \emph{wikipedia.org}, while \bcp returns multiple domains, among others \emph{gettyimages.com}.


Focusing on the link's relevance, 
\tabref{tab:related_to_queries} shows the percentage of the links that (i)~are related to the query (Yes),
(ii)~are not related to the query (No), (iii)~are related to the query, but with a different programming language,
and (iv)~are not reachable or do not have code snippets at all (NA). 
For \bcp, we obtained higher percentages of relevant links 
(41.03\% for Ruby up to 61.79\% for Java vs 24.26\% for Python up to 56\% for Go). 
The percentage of links related to the query, yet containing snippets in a different programming language, tends to be greater for \bcp than for \gem, although we do not know their implementation details, and we cannot tell whether one is better than the other for providing attribution.

As an example, when asking for a Java snippet to compute the \textit{confusion matrix}, \bcp returned three links: one pointing to a related Stack Overflow post (\# 39657396)\footnote{To access the link use \url{https://stackoverflow.com/questions/ID} replacing ID with the question identifier reported in the text.}, and the other two explaining what a confusion matrix is and how to compute it with Python~\cite{tutorialconf,officialconf}. This may be due to the widespread usage of Python for data-intensive/Machine Learning (ML) software.
When we asked \bcp to provide a Python snippet for the same task, all the five links provided
were related to the query.

\begin{table}[t!]
\caption{Descriptive statistics of the Jaccard metrics between links provided by LLM-based assistants and search engines}
\label{tab:jaccard}
\centering
\small
\resizebox{0.85\columnwidth}{!}{
\begin{tabular}{lrrrrrr}
    \multicolumn{7}{c}{\sc \bcp} \\ 
    \midrule
    \textbf{Languages} & \textbf{Min} & \textbf{1Q} & \textbf{Median} & \textbf{Mean} & \textbf{3Q} & \textbf{Max} \\
    \midrule
      \textbf{Go} & 0.11 & 0.14 & 0.20 & 0.26 & 0.30 & 0.60 \\ 
      \textbf{Java} & 0.00 & 0.08 & 0.20 & 0.23 & 0.33 & 0.67 \\ 
      \textbf{JavaScript} & 0.00 & 0.08 & 0.18 & 0.19 & 0.29 & 0.67 \\ 
      \textbf{PHP} & 0.00 & 0.12 & 0.17 & 0.22 & 0.25 & 0.60 \\ 
      \textbf{Python} & 0.00 & 0.00 & 0.17 & 0.19 & 0.26 & 1.00 \\ 
      \textbf{Ruby} & 0.00 & 0.10 & 0.18 & 0.16 & 0.20 & 0.33 \\ 
   \midrule 
   \multicolumn{7}{c}{\sc \gem} \\ 
   \midrule
   \textbf{Languages} & \textbf{Min} & \textbf{1Q} & \textbf{Median} & \textbf{Mean} & \textbf{3Q} & \textbf{Max} \\
   \midrule
\textbf{Go} & 0.00 & 0.00 & 0.00 & 0.03 & 0.00 & 0.20 \\ 
  \textbf{Java} & 0.00 & 0.00 & 0.00 & 0.04 & 0.00 & 0.33 \\ 
  \textbf{JavaScript} & 0.00 & 0.00 & 0.00 & 0.04 & 0.00 & 0.33 \\ 
  \textbf{PHP} & 0.00 & 0.00 & 0.00 & 0.04 & 0.00 & 0.33 \\ 
  \textbf{Python} & 0.00 & 0.00 & 0.00 & 0.04 & 0.00 & 0.33 \\ 
  \textbf{Ruby} & 0.00 & 0.00 & 0.00 & 0.02 & 0.00 & 0.20 \\ 
   \bottomrule
\end{tabular}
}
\end{table}

As a last analysis, we discuss the overlap between the links provided by the two LLM-based assistants and those retrieved from the search engines for the same query. \tabref{tab:jaccard} reports descriptive statistics of the obtained distribution of Jaccard overlap~\cite{jaccard}.
For \gem, there are almost no links in common with those retrieved by the search engine, with the median value (together with the third quartile) being 0 on the whole set of programming languages being considered. The median value increases up to 0.2 for Java and Go when considering \bcp, with a third quartile being never null across all considered programming languages. 
As an example, when asking \bcp to provide a Java snippet to \textit{extract data from html content}, we received only one relevant link, yet related to JavaScript (\# 28899298).
In contrast, Bing retrieved at least two relevant documents that rely on Java, \ie a tutorial~\cite{tutorialhtml} and a Stack Overflow post (\#~7360411). A different worth-mentioning example is the one in which we asked \bcp to provide a Python snippet to \textit{convert json to csv}. In this case, the four links provided by the assistant did not contain any Python code, \ie they were sandboxes. When we queried Bing for the same task, the top four links pointed to tutorials~\cite{tutorialjson2,tutorialjson1,tutorialjson3} and a Stack Overflow post (\# 1871524) related to the submitted query. These examples indicate that links provided by the assistants should be used with caution, as they may be incomplete or unrelated to the query.
We are unaware of how LLM-based assistants perform Web searches. Given the low overlap we found, they might not send the query to the search engines directly, but rather, they may leverage direct access to their internal representation. However, they are also likely using queries that are different from ours.

\begin{resultbox}
\textbf{RQ$_1$ summary:} 
LLM-based assistants output several, heterogeneous, and noisy links for a query.  
\bcp and \gem differ for the number and type of links provided.
Finally, LLM-based assistants' link retrieval is not directly reproducible by simply querying their related search engines. 
\end{resultbox}

\begin{table}[t]
\caption{Links related to the generated snippets (according to manual annotation)}
\label{tab:related_to_snippets}
\centering
\small
\resizebox{0.8\columnwidth}{!}{
\begin{tabular}{lrrrr}
  \multicolumn{5}{c}{\sc \bcp} \\ 
  \midrule
  \multirow{2}{*}{\textbf{Languages}} & \multirow{2}{*}{\textbf{No}} & \multicolumn{3}{c}{\textbf{Yes}} \\ 
  \cmidrule{3-5}
   &  & \textbf{One} & \textbf{$>$ One} & \textbf{Tot.}  \\ 
  \midrule
    \textbf{Go} & 28.57\%  & 28.57\% & 42.86\% & 71.43\% \\ 
    \textbf{Java} & 52.17\%  & 34.78\% & 13.04\% & 47.83\% \\ 
    \textbf{JavaScript} & 28.57\% & 46.43\% & 25.00\%  & 71.43\% \\ 
    \textbf{PHP} & 16.00\%  & 44.00\% & 40.00\% & 84.00\% \\ 
    \textbf{Python} & 29.09\%  & 43.64\% & 27.27\% & 70.91\% \\ 
    \textbf{Ruby} & 42.86\%  & 35.71\% & 21.43\% & 57.14\% \\ 
    \midrule 
  \multicolumn{5}{c}{\sc \gem} \\ 
  \midrule
  \multirow{2}{*}{\textbf{Languages}} & \multirow{2}{*}{\textbf{No}} & \multicolumn{3}{c}{\textbf{Yes}} \\ 
  \cmidrule{3-5}
   &  & \textbf{One} & \textbf{$>$ One} & \textbf{Tot.}  \\ 
  \midrule
      \textbf{Go} & 85.71\%  & 14.29\% & 0.00\% & 14.29\% \\ 
      \textbf{Java} & 93.75\%  & 6.25\% & 0.00\%  & 6.25\%\\ 
      \textbf{JavaScript} & 90.00\%  & 10.00\% & 0.00\% & 10.00\% \\ 
      \textbf{PHP} & 96.00\%  & 0.00\% & 4.00\%  & 4.00\%\\ 
      \textbf{Python} & 93.75\%  & 4.17\% & 2.08\%  & 6.25\%\\ 
      \textbf{Ruby} & 100.00\%  & 0.00\% & 0.00\%  & 0.00\%\\ 
  \bottomrule
\end{tabular}
}
\end{table}

\subsection{RQ$_2$: \rqtwo}
\tabref{tab:related_to_snippets} reports, for the different programming languages, whether or not the links provided by the LLM-based assistants contain---according to our manual assessment---at least one similar snippet. Specifically, we report the percentage of cases with no relevant links, exactly one link, more than one link (likely implying ambiguity in terms of attribution), and at least one relevant link (indicating the possibility of verifying trustworthiness and providing attribution). As explained in \secref{sec:manual}, for Go and Ruby results concern a limited number of snippets.

\begin{table}[t]
\caption{Number of links from different link types containing snippets relevant to the ones generated by the LLM-based assistants}
\label{tab:relevantLinks_bySource}
\small
\centering
\resizebox{\columnwidth}{!}{
\begin{tabular}{lrrrrrr}
    \multicolumn{6}{c}{\sc Bing CoPilot} \\ 
    \midrule
    \multirow{2}{*}{\textbf{Languages}} & \multirow{2}{*}{\textbf{Q\&A}} & \multirow{2}{*}{\textbf{Repository}} & \multirow{2}{*}{\textbf{Tutorial}} & \textbf{Official} & \textbf{Gist/Code} \\ 
    & & & & \textbf{Doc.} & \textbf{Example}\\
    \midrule
        \textbf{Go} &   3 &   2 &   3 &   0 &   2  \\ 
        \textbf{Java} &   9 &   9 &  14 &   0 &   0 \\ 
        \textbf{JavaScript} &  15 &  19 &  11 &   1 &   0  \\ 
        \textbf{PHP} &  15 &  19 &   8 &   2 &   1 \\ 
        \textbf{Python} &  26 &   3 &  36 &   2 &   2 \\ 
        \textbf{Ruby} &   8 &   1 &   0 &   2 &   0  \\ 
    \midrule 
    \multicolumn{6}{c}{\sc Gemini} \\ 
    \midrule
    \multirow{2}{*}{\textbf{Languages}} & \multirow{2}{*}{\textbf{Q\&A}} & \multirow{2}{*}{\textbf{Repository}} & \multirow{2}{*}{\textbf{Tutorial}} & \textbf{Official} & \textbf{Gist/Code} \\ 
    & & & & \textbf{Doc.} & \textbf{Example} \\
    \midrule
        \textbf{Go} &   1 &   0 &   0 &   0 &   0 \\ 
        \textbf{Java} &   2 &   0 &   1 &   0 &   0 \\ 
        \textbf{JavaScript} &   1 &   0 &   2 &   0 &   0 \\ 
        \textbf{PHP} &   1 &   0 &   1 &   0 &   0 \\ 
        \textbf{Python} &   0 &   0 &   2 &   3 &   0 \\ 
        \textbf{Ruby} &   0 &   0 &   0 &   0 &   0 \\ 
    \bottomrule
\end{tabular}}
\vspace{-2mm}
\end{table}

By looking at \tabref{tab:related_to_snippets}, it is possible to notice the different behavior among \bcp and \gem, with \gem rarely providing a relevant link, \ie percentages below 15\% (0\% for Ruby). Therefore, we will primarily focus on discussing the results from \bcp. For \bcp, the percentage of queries with at least one relevant link available for checking trustworthiness or providing attribution ranges from 47.83\% (Java) to 84\% (PHP). We can also observe a relatively high percentage of cases where there is more than one relevant link for the same query (40\% for PHP and 42.86\% for Go). 
Even though the results for \bcp appear to be good, there are still instances where the LLMs struggle to ascertain ``where the snippet is likely coming from''. In all those cases, it is up to the developer to determine the likely provenance of the generated snippet.  As an example, when querying \bcp to \textit{encrypt aes ctr mode} in PHP, we received eight links. Two of these links contain snippets similar to the generated one, namely a Stack Overflow post (\# 3422759) and a file from a GitHub project~\cite{githubencrypt}, indicating these links could have been used as sources from which the LLM has generated its code. This is supported by the quantitative similarity analysis, which shows at least a 64.86\% cloning ratio and a textual similarity of 0.94. However, such similarity is only an indicator, and the burden of determining whether the automatically generated code originates from the provided link is still up to the developers. This can be done, for instance, by comparing the date of a post on Stack Overflow or the date when the similar code was pushed to a GitHub repository.

Sometimes, the automatically generated snippet contains a ``fingerprint" that relates it to a specific Web domain. As an example, when querying \bcp to \textit{output to html file} using Python, the provided snippet included the statement: \texttt{URL = 'https://www.geeksforgeeks.org/how-to-\\scrape-all-pdf-files-in-a-website/'}, referring to \textit{geeksforgeeks.org}. Among the provided links, we found one~\cite{gfghtml} pointing to this domain containing a snippet identical (Type 1 clone) to the generated one. 

\begin{figure}[t!]
    \centering
    \includegraphics[width=1\linewidth]{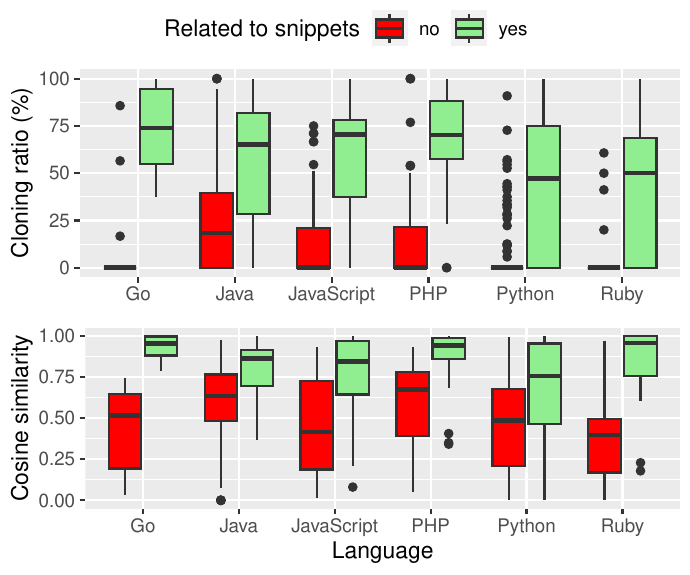}
    \caption{\bcp: Cloning ratio and textual similarity}
    \label{fig:cloneCoPilot}
    \vspace{-2mm}
\end{figure}

\begin{figure}[t!]
    \centering
    \includegraphics[width=1\linewidth]{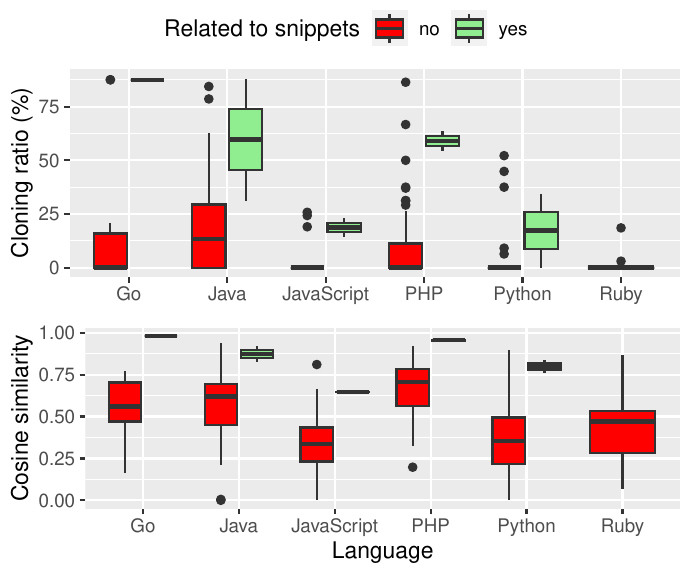}
    \caption{\gem: Cloning ratio and textual similarity}
    \label{fig:cloneGemini}
    \vspace{-2mm}
\end{figure}

\tabref{tab:relevantLinks_bySource} deepens our analysis by reporting the number of links of different types, providing code snippets related to the generated ones. We omit source types for which no relevant link was provided. Again, we only discuss results for \bcp as the figures for \gem are extremely low, mainly due to the low number of queries with at least one relevant link being provided. 
Looking at the results, JavaScript and PHP show a prevalence of relevant links from GitHub, Q\&A and tutorials. For Java and Python, we found a prevalence of tutorials and, secondarily, Q\&A forums. Q\&A forums are also the main type for Ruby, while the (few) cases for Go are spread across different types. 
As the table shows, we rarely found a match on the official documentation---which is mostly syntax explanation and rarely provides complete examples---and on code example websites (\eg Gist).

As mentioned, we leverage clone detection and textual similarity as a complement to the performed manual assessment. Results are reported in \figref{fig:cloneCoPilot} and \figref{fig:cloneGemini}.
First, through Spearman's rank correlation, we found that the cloning ratio and textual similarity information are consistent ($\rho=0.76$ for \bcp and $\rho=0.59$ for \gem). 
For \bcp, we found that the maximum cloning ratio and the textual similarity computed across snippets from links classified as ``not relevant" and ``relevant" are statistically significant (Wilcoxon rank sum $p$-value $<0.001$ in both cases) with a (large) effect size of $d=0.55$ and $d=0.60$, respectively.  Differences are statistically significant also for \gem, ($p$-value $<0.001$ in both cases), with large effect sizes, $d=0.79$ and $d=0.83$ for cloning ratio and textual similarity.

Looking deeper at the \bcp cloning ratio plots, we can see how the median cloning ratio for the ``yes'' is above 50\% for all programming languages but Python and Ruby, where boxes show a large variability. 
We can notice that Python and Ruby ``yes'' boxes have a first quartile at zero. As we checked manually, this is due to the minimum clone threshold (of 20 tokens). Some relevant Python or Ruby snippets, thanks to the heavy usage of libraries, turn out to be relevant even if they are relatively smaller. Also, sometimes, a code example (returned as a whole by the LLM-based assistants) is piecewise explained as multiple snippets in tutorials. This makes the cloning fail. As an example, when asking \bcp to provide a Java code snippet about \textit{all permutations of a list}, one of the external links~\cite{perm} from which the automatically generated code may originate contains the snippet, but spread across multiple ones. For this reason, the maximum cloning ratio and the maximum textual similarity are very low, 0.0\% and 0.36, respectively.

The other interesting case is the ``no'' with a very high cloning ratio (notice some outlier points for most of the languages). As an example, when asking \bcp to provide a Java code snippet about \textit{regex case insensitive}, one of the provided links~\cite{regex} contains a code snippet with a 100\% cloning ratio and a very high textual similarity (0.83). However, the annotators categorized the link as ``not relevant to the snippet.'' While the code structure between the generated code snippet and the one in the link was very similar (hence 100\% cloning ratio for type-2 clones), the code snippets contained some different identifiers, literal values, and comments.

For \gem (note that, for Ruby, the figure only shows the ``no'' box as we found no relevant links), results are less interesting to discuss due to the fairly limited number of ``yes'' cases.
One can notice a smaller median cloning ratio of both JavaScript and Python for the ``yes'' links, and a smaller variability (due to the fewer cases). Nevertheless, when looking at the textual similarity, median values for the ``yes'' still tend to be on the high side, above 0.50 for JavaScript and above 0.75 in all other cases.

\begin{resultbox}
\textbf{RQ$_2$ summary:} \bcp is able to provide at least one ``relevant'' link for the majority of the considered queries, 
The relevant link types are mostly repositories, tutorials, and Q\&A. Furthermore,  the results of the clone detection and textual similarity analysis seem to be consistent with the results of the manual assessment.
\end{resultbox}

%% file: implications.tex
Based on the results detailed in Section~\secref{sec:results}, we distilled the following key findings.
 
\textbf{LLM provide links varying in numbers and nature, also depending on the programming language being considered.
} In our assessment, we already discovered how the direct use of LLMs did not produce relevant information. Indeed, the only way for DL models to produce attribution would be to specifically train them with links (direct model-driven attribution) which, as explained in the introduction, is discouraged because it may lead to hallucinations \cite{li2023surveylargelanguagemodels}.
Instead, the investigated LLM-based assistants---\bcp and \gem---provide links in different numbers, nature, and ability to match the generated code. 

\textbf{Links are unlikely to originate from a straightforward Web search.} While we do not know the exact reasons, we conjecture that \bcp and \gem exploit direct access to the Bing and Google search engine knowledge base, as well as specific, proprietary querying strategies used to retrieve links. This means that developers may struggle to obtain the same results by replicating the LLM query to the search engine. This may diminish the trustworthiness of the obtained links.

\textbf{The list of provided links can be long.} To find a likely link from which the automatically generated code snippet may originate, users have to scrutinize several links, many of which are irrelevant, as being invalid, unrelated to the query, related to the query but with different programming languages, or being generic documentation of APIs and pieces of technology. 
This may suggest that development tools could augment the LLM assistant output with some more advanced provenance analysis (\eg using similarity or clone detection) to highlight the most relevant links.

\textbf{Limited, and no direct provenance information.} Two studied assistants provide links as ``a way one could deepen knowledge about the query topic,'' which means learning about the language, technology, or, in the best case, about possible ways to implement the coding task. 
Such links do not have the purpose of telling which source should be considered to evaluate the snippet's level of trust, attribute ownership, or consider to determine licensing terms. Moreover, there are many cases in which none of the produced links contain code snippets having a high likelihood of being the origin of the generated ones. Furthermore, sometimes there is no one-to-one match, as a code example could be discussed piecewise in a tutorial.  This may possibly suggest that such a snippet could also be present (entirely, this time) somewhere else on the Web, yet the LLM-based assistants did not provide its link. 

\textbf{LLMs still produce verbatim copies from likely training sets.} This is not a new finding, but, rather, a confirmation of what was found by previous studies \cite{Chu_Song_Yang_2024,karamolegkou-etal-2023-copyright,KatzyPDI24,longpre2023dataprovenanceinitiativelarge,ZhangXL0HXS0024}. However, such a finding, along with the lack of precise provenance information, suggests that, nowadays, a trustworthy, licensing-aware usage of LLM-generated code is still quite problematic and immature.

Our study allowed us to trigger the following implications for practitioners, researchers, and LLM providers.

\textbf{Practitioners} have, with the currently available technology, still a hard life. On the one hand, it can be difficult to determine if the generated code can be trusted. On the other hand, our findings show that its inappropriate reuse can lead to licensing inconsistencies or, ultimately, legal implications. 

\textbf{Software Engineering Researchers}, while using LLMs and related tools as black boxes (or almost), could explore the implementation of tools to support developers. This may be done, for example, by automating the analysis performed in our study---\ie determining the link with the most related snippet(s) and providing developers with licensing, authorship, and source reputation information.

\textbf{LLM providers}, while our work is scoped within software engineering, could be interested in setting up mechanisms to provide provenance information to the LLMs' users. It could be part of the training process or implemented by combining the LLM with some Web (or training set, if available) search. The latter needs to be done starting from the generated code snippets, rather than from the query.


%% file: threats.tex
Threats to \emph{construct validity} concern the relationship between theory and observation. The most important threat is related to determining the extent to which a given online snippet can be a likely provenance for the LLM-generated one.  Determining whether two snippets are clones is a challenging task for humans~\cite{CharpentierFMYR17}. To mitigate this threat, as explained in \secref{sec:manual}, we established a shared set of criteria, leveraged multiple annotators assessing the same links, and complemented the manual analysis with an automated analysis based on clone detection and textual similarity. 

Threats to \emph{internal validity} concern factors internal to the study that could impact our results.
A first threat can be related to the choice of prompts and interactions. While we aligned them to LLM guidelines and previous research, and tried different options, we cannot exclude that there could be prompts providing different results. Nevertheless, we noticed that prompts might influence the LLMs' ability to produce valid/useful results in terms of snippets, rather than in providing relevant links.
Beyond what we discussed in \secref{sec:querying}, we performed a small analysis to check whether a more complex approach, \ie using a chain-of-thought, could have produced better results. We conducted this analysis on 30 random queries, five for each programming language, for each LLM-based assistant. After obtaining the list of links, we asked ``Which of these sources has the most similar code to the one you generated above?'' Through manual validation, we found that for \bcp, none of the links the LLM indicated as the one with the most similar content was relevant, while for \gem this happened in four cases. Hence, while a chain of thought could be valuable, it is still insufficient for retrieving better information.

\revised{LLMs could suffer from non-determinism, also considering that the used web-based LLM assistants do not provide advanced hyperparameter settings. Given the nature of this study (requiring a manual analysis) we could not afford to execute the queries multiple times. We mitigate this threat by reporting statistics on a relatively large number of query results featuring 1,520 links to be manually validated.} 

Another threat to internal validity is related to human annotators' error-proneness and subjectivity, mitigated by having multiple annotators and computing inter-rater agreement (Cohen's $k$~\cite{cohen1960coefficient}).
A further threat to the internal validity could be related to cloning ``by chance,'' which has been mitigated by filtering out small snippets and by manually excluding queries for which all the automatically generated code snippets were considered ``trivial''.

Finally, although this is less likely, and although some domains such as Stack Overflow forbid posting AI-generated code~\cite{soAIpolicy}, we cannot prove the direction of the code reuse, \ie we cannot exclude that the code on the provided links was produced by the LLMs. Through a manual check, we found that 28 sources (mainly tutorials) found by \bcp and 1 found by \gem were updated in a date between the latest checkpoints of the LLMs and the date of our analyses.

Threats to \emph{external validity} concern the generalizability of our findings. For practical reasons, this study focused on only two LLM-based assistants, \ie \bcp and \gem. \revised{It must be said that \bcp is based on GPT-4, the same LLM leveraged by other tools used in software development, such as ChatGPT~\cite{chatgpt} and GitHub Copilot Chat~\cite{github_copilot}.}
As the field evolves rapidly, the reliability of AI systems may change, and new/updated LLMs could offer improved capabilities.
A further threat concerns the choice of the LLM queries. Although we leveraged a widely adopted benchmark~\cite{husain2020codesearchnetchallengeevaluatingstate}, results could be different with other queries, \eg those related to more complex tasks and requiring specific technology. 
Another threat to external validity concerns the studied programming languages. While we considered six of the top popular programming languages in the open-source community, the results may not be generalized to the universe of programming languages.

\emph{Conclusion validity} threats are related to the relationship between observation and outcome. This is merely an exploratory study, answered through descriptive statistics and by discussing examples. Furthermore, where appropriate (automated similarity analysis), we leveraged suitable statistical procedures.

%% file: related.tex
In this section, we discuss related work about (i)  code provenance analysis, (ii) code search, and (iii) legal concerns with AI-generated code.

\subsection{Approaches and Studies About Code Provenance} 
The software engineering literature has reported various approaches (and related studies), relying on code hashing, clone detection or code fingerprinting, to establish the provenance of a given source code artifact, \ie to determine where it was likely copied from. 

German \etal~\cite{GermanPGA09} conducted a study combining clone detection with data from versioning systems to trace the migration of source code files across Unix kernels. Differently, Rousseau \etal~\cite{RousseauCZ20}, relied on artifact hashing to study code provenance. Specifically, they investigated how artifacts are replicated across multiple repositories, whether this depends on the presence of forked repositories, and how the growth of original files evolves. A limitation of the hashing-based approach is the fixed granularity of the considered artifacts (\eg file level), which cannot be used in our context. 

Another way of identifying the provenance of software artifacts---to some extent similar to metric-based clone detection~\cite{MayrandLM96}---is software bertillonage. It takes the name from Alphonse Bertillon \cite{bertillon}, the inventor of a method for identifying criminals through physical measurements. Like people identification, software bertillonage~\cite{DaviesGGH13} characterizes a software artifact with some of its metrics.  The study by Davies \etal~\cite{DaviesGGH13} empirically identified the provenance of Java classes on the Maven Central Repository by leveraging the class ``anchored class signature,'' featuring information such as name, inheritance tree, fields, methods, and others. 
A related approach has been used by Zacchiroli \etal~\cite{SunGZ23} in the context of Python packages.
They analyzed the uniqueness of global identifiers, finding that only 76\% of identifiers are unique to a single package. 

Last but not least, other initiatives related to software provenance include those of companies such as Blackduck~\cite{blackduck,blackduck2}, as well as open-source tools such as FoSSology~\cite{Gobeille08}, which also perform licensing analysis.

\subsection{Code Search Approaches} 
Code search is a core software engineering task aiming to find relevant code snippets from a code base~\cite{Liu2021,Sun2024}. Semantic code search approaches retrieve relevant code given a natural language query, searching code snippets based on their meaning rather than based on keyword or syntax similarity, \eg \cite{gallardo2009internet,phan2024leveraging,ReissICSE2009,Stolee2014}. Early code search tools were based on information retrieval techniques since keywords from comments and variable names were often sufficient for finding reusable code~\cite{ReissICSE2009}. Other work attempted to improve results based on refinement of queries, code base, and/or matching between queries and code snippets~\cite{Sun2024,zhang2017expanding}. 

McMillan~\cite{Mcmillan2013} proposed Portfolio, a graph-based code search engine, and showed how it outperforms commercial products such as Google Code Search and Koders.


Stolee~\etal~\cite{Stolee2014}, instead, leveraged Statistical Machine Translation solvers to translate program specifications into constraints, allowing for a code retrieval that satisfies programmers' requirements. Although their approach has limitations in terms of flexibility and efficiency, it can be used in various programming languages.

Gu~\etal\cite{Gu2018} proposed the CODEnn approach. It jointly embeds code snippets and natural language descriptions into a high-dimensional vector space. 
Shuai \etal~\cite{Shuai2020} overcame the semantic mismatch limitations of CODEnn, by proposing CARLCS-CNN7, which incorporates a CNN with a co-attention mechanism to learn better representations. 
Phan and Jannesari~\cite{phan2024leveraging} proposed Oracle4CS, an approach leveraging Statistical Machine Translation to improve code search.

Further details about code search approaches can be found in secondary studies by Liu~\etal~\cite{Liu2021} and by Sun~\etal~\cite{Sun2024}.

We do not focus on code search directly, yet we claim that code search can be a key component to support provenance identification, given an LLM-generated code snippet.


\subsection{Legal Issues of LLMs Generated Code}


There could be legal consequences from reusing and redistributing software artifacts~\cite{VendomeGPBVP18,Wolter2023}, and this is even more critical for AI-generated ones~\cite{Chu_Song_Yang_2024,longpre2023dataprovenanceinitiativelarge,ZhangXL0HXS0024}.  This is because we do not know the terms under which we can reuse LLMs generated code, as it could be derived (through training) from unknown software artifacts taken from the Web~\cite{sun2022coprotector}. That is, generative models trained on large data corpora can generate outputs being verbatim copies or at least derivative work of copyrighted text/code used for their training~\cite{Chu_Song_Yang_2024}.

Karamolegkou \etal~\cite{karamolegkou-etal-2023-copyright} empirically investigated copyright violations of LLMs in popular books and coding problems. They experimented with some (open- and closed-source) LLMs and probing methods.
Their findings suggest that LLMs memorize and reproduce copyrighted text.
To some extent, this study is the one closest to ours. However, our aim is not to show that LLMs violate copyright but, rather, to determine whether, through links provided by LLM-based assistants, it is possible to identify the (likely) provenance of the generated code snippets.


Katzy~\etal~\cite{KatzyPDI24} studied the current trends in the code license infringements in the LLMs training dataset and the importance of incorporating code into the training of large language models. Their results indicated that every dataset contained license inconsistencies, despite being selected based on their associated repository licenses.

Majdinasab~\etal~\cite{abs-2402-09299} proposed TraWiC, a model-agnostic, interpretable method for detecting code inclusion in an LLM’s training dataset. TraWiC extracts syntactic and semantic identifiers to train a classifier for detecting code inclusion. 

Our work complements the aforementioned studies on LLM copyright infringements. This is because if AI-based assistants could provide suitable provenance information for the recommended code, developers would be able to provide proper attribution and licensing.


%% file: conclusion.tex
In this paper, we present the findings of an empirical study investigating the extent to which LLM-based assistants---and specifically \bcp and \gem---produce links to aid in determining the provenance of automatically generated code. We leveraged queries from \textsc{CodeSearchNet}~\cite{husain2020codesearchnetchallengeevaluatingstate} and instantiated on \lang programming languages, combined with a manual assessment of the LLM-based assistants' output with an automated one based on clone detection and textual similarity.

The obtained results indicate 
(i)~a general noise of the produced links, and (iii)~ a limited ability of the LLM-based assistants to provide valuable provenance information. 

\revised{
Redistributing derivative work infringing on licensing or attribution constraints may lead to legal consequences and can be problematic for AI-generated code. This is because one does not know the terms under which LLM-generated code can be reused, as it could be verbatim copies or derivative work of copyrighted text/code taken from the Web for training. Our study is a first step in evaluating the capability of the LLMs to provide links to web resources containing code snippets highly similar/relevant to the ones generated by the LLM.
}

As the subject of our research field rapidly evolves, future work should replicate this study using emerging technologies (\eg ChatGPT~\cite{chatgpt} when utilized with GPT-4o). Moreover, our findings open the road for better LLMs and also LLM-based assistants able to provide better support for the likely provenance of the generated software artifacts.